\documentclass{an}
\usepackage{graphicx}
\usepackage{times}
\usepackage{fancyhdr}
\newcommand{\kms}{\,{\rm km\,s^{-1}}}

\newcommand{\kpc}{\,{\rm kpc}}
\newcommand{\Mpc}{\,{\rm Mpc}}

\newcommand{\mkG}{\,\mu{\rm G}}

\newcommand{\radm}{\,{\rm rad\,m^{-2}}}

\newcommand{\ini}{{}_\mathrm{i}}

\newcommand{\Rey}{\mathrm{Re}}
\newcommand{\Reycr}{\mathrm{Re}_\mathrm{cr}}
\newcommand{\Rm}{R_\mathrm{m}}
\newcommand{\Rmcr}{R_\mathrm{m,cr}}

\newcommand{\RM}{\mathrm{RM}}
\newcommand\sfrac[2]{{\textstyle{\frac{#1}{#2}}}}

\newcommand{\phzero}{\phantom{0}}
\newcommand{\tphzero}{\phantom{0}\phantom{0}}

\begin{document}

\title{The origin and evolution of cluster magnetism}

\author{
A.~Shukurov\inst{1,2}
\and
K.~Subramanian\inst{2,1}
\and
N.~E.~L.~Haugen\inst{3}
}

\institute{
$^1$School of Mathematics and Statistics, University of Newcastle, Newcastle
        upon Tyne, NE1 7RU, U.K.\\
$^{2}$Inter-University Centre for Astronomy and
        Astrophysics,  Post Bag 4, Ganeshkhind, Pune 411 007, India\\
$^3$Department of Physics, Norwegian University of Science and Technology,
        H{\o}yskoleringen 5, 7034 Trondheim, Norway
}

\date{Received  2006 Mar 02; accepted 2006 Mar 21; published online 2006 May 11}
\lhead{A.~Shukurov, K.~Subramanian, N.~E.~L.~Haugen}
\rhead{The origin and evolution of cluster magnetism}

\abstract{Random motions can occur in the intergalactic gas of galaxy
clusters at all stages of their evolution. Depending on the poorly known
value of the Reynolds number, these motions can or cannot become turbulent,
but in any case they can generate random magnetic fields via dynamo
action. We argue that magnetic fields inferred observationally for the
intra\-cluster medium require dynamo action, and then
estimate parameters of random flows and magnetic fields at various
stages of the cluster evolution.
Polarization in cluster radio halos predicted by the model would be
detectable
with the SKA.
\keywords{ Galaxies: clusters: general --  galaxies: intergalactic medium
-- galaxies: magnetic fields -- MHD -- turbulence}}

\correspondence{anvar.shukurov@ncl.ac.uk (AS)}
%\correspondence{anvar.shukurov@ncl.ac.uk (AS); kandu@iucaa.ernet.in (KS);
%nils.haugen@phys.ntnu.no (NELH)}

\maketitle

%--------------------------------------------------------------------------
\begin{figure*}
\begin{center}
\resizebox{0.75\hsize}{!}
{\includegraphics[width=0.48\textwidth,clip]{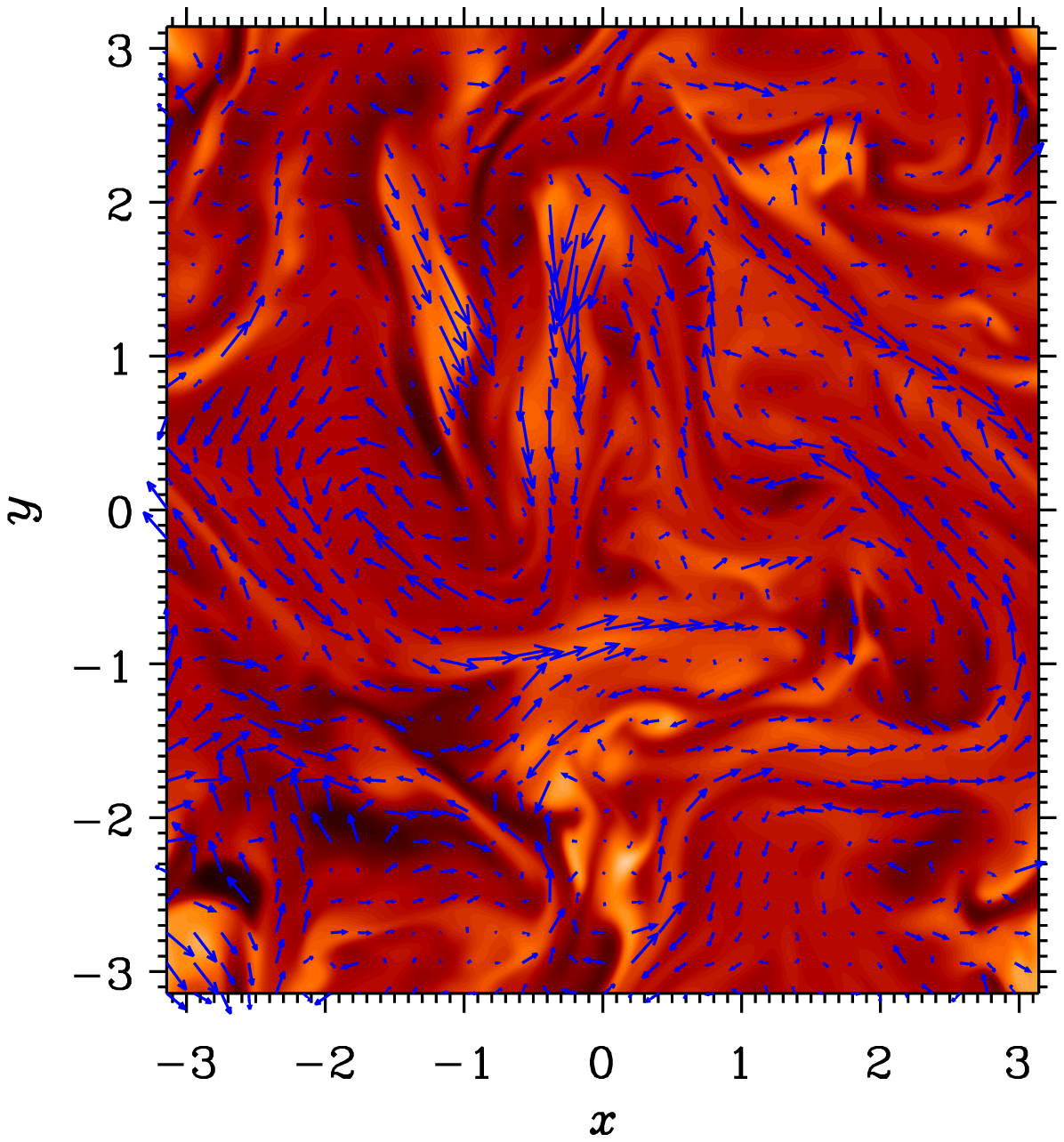}
$\qquad$%\hfill
\includegraphics[width=0.49\textwidth,clip]{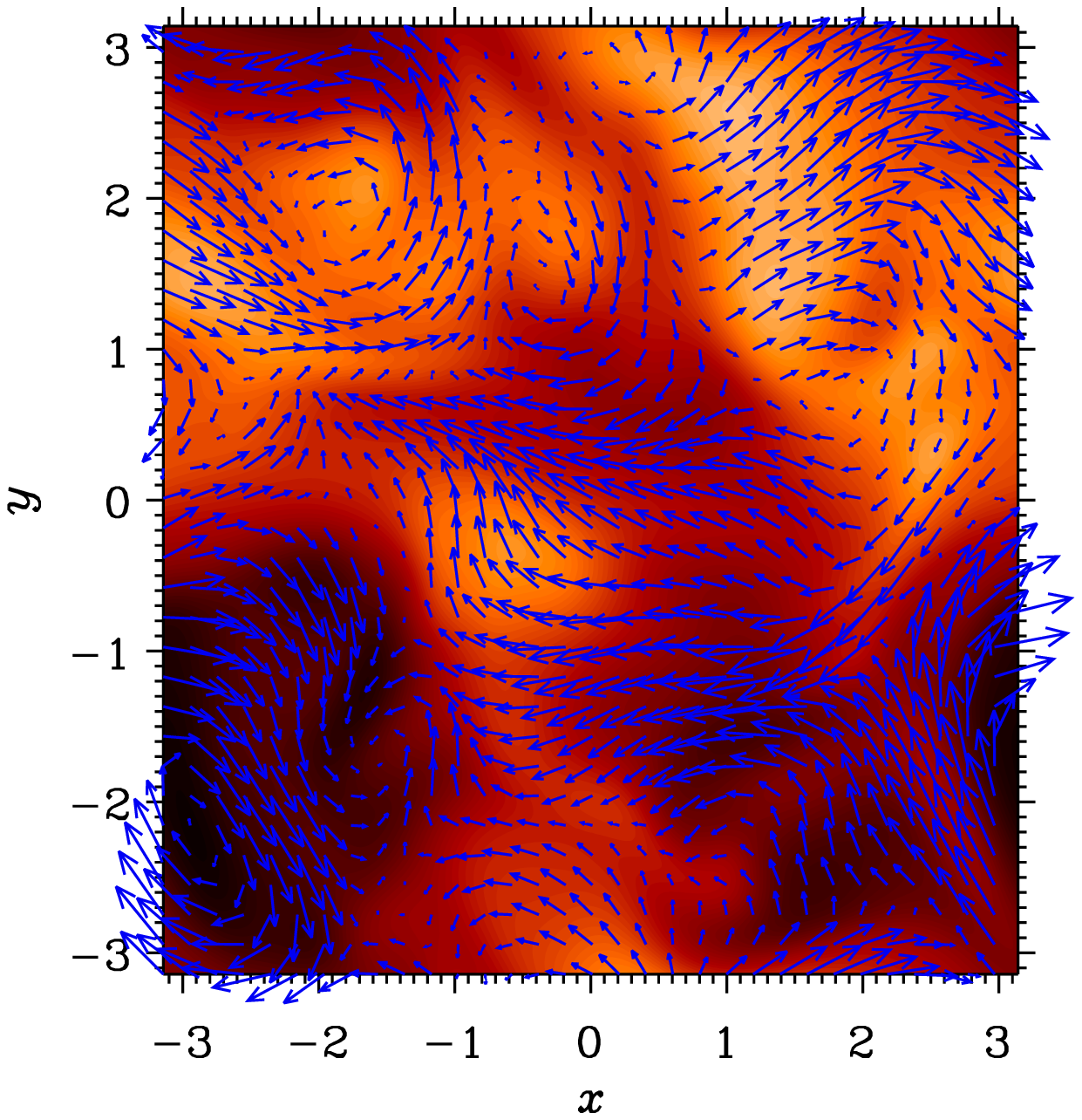}}
\end{center}
\caption{Snapshots of magnetic field in a cross-section through the middle of the
computational domain
in a numerical simulation of turbulence driven by an imposed random force
and its dynamo action. The left-hand panel shows a statistically steady
state at a time $t/t_0\ini=0.30$
whereas the right-hand panel illustrates magnetic field structures in
turbulence at a late stage of decay, $t/t_0\ini\approx60$.
Here $t_0\ini$ is the
eddy turnover time before the start of the turbulence decay
(given in Table~\ref{res}). The dimensionless energy injection scale in these
simulations is about 4 (with the domain size of $2\pi$),
so each frame contains a few turbulent cells.
The strongest magnetic field within the frame is close to the equipartition
value with respect to the turbulent energy.
The magnitude of the field component perpendicular to the
plane of the figure is shown color coded (in shades of grey) with black
corresponding to field pointing into the
figure plane, and
lighter shades, to field pointing out of the plane. The field in the plane of
the figure is shown with vectors whose length is proportional to the field
strength.}
\label{snapshots}
\end{figure*}
%--------------------------------------------------------------------------

%-------------------------------------------------------------------------
\section{Introduction}
We witness the epoch when galaxy clusters are being formed.
Galaxies and smaller-size structures have generally achieved (quasi-)steady
states in their evolution, whereas structures
of larger scales keep evolving.
There is abundant evidence of formation processes in galaxy clusters, including
the merger of large
masses comparable to the cluster
mass. Relaxed, symmetric galaxy clusters are rare.
Thus, most constituents of the clusters keep
evolving. Our particular interest here is with motions of the gas and magnetic
fields in the intergalactic space of evolving galaxy clusters.
The epoch of major mergers is notable for
widespread, intense random motions, followed by a period of decaying flows. In
a steady state, random motions can be confined to the wakes produced
by the cluster galaxies and smaller mass subcluster
clumps that continue falling into the cluster;
as we argue here, these motions
are likely to occupy a small fraction of the total volume.

The high electrical conductivity of the intra\-cluster gas is not
sufficient to
ensure the survival of a magnetic field captured by the forming cluster.
The reason is that the intra\-cluster gas is very
viscous, and any inhomogeneous magnetic field would drive gas motions which will
rapidly decay, thereby converting magnetic energy into heat.
Even for a small viscosity, the field would decay by driving
decaying MHD turbulence. On the other hand, random gas motions
of the intra\-cluster gas can be efficient generators
of (random) magnetic fields via a process known as the fluctuation dynamo.

The origin and properties of magnetic field in the intra\-cluster plasma cannot
be understood without knowing the nature and parameters of the plasma
motions. The reason is that the magnetic Reynolds number in the intra\-cluster
gas is large, hence magnetic induction effects are strong for any plausible
speed of gas motion.
Our model of gas motions and magnetic fields in galaxy clusters,
presented and justified in detail by Subramanian et al.\ (2006),
is in close agreement with the available data.
It predicts that
magnetic field in the intra\-cluster gas is represented by magnetic sheets of a
thickness 20--30 kpc, with magnetic field strength about $2$--$4\mkG$ within the
sheets. These structures fill about 20\% of the volume of a single
turbulent cell. Apart from this intermittent component, magnetic field
also has a weaker, widely distributed part.

We propose that, at late stages of cluster evolution, turbulence is confined
to the wakes of galaxies and infalling mass clumps, and does not fill the
volume. However, the area covering factor of the wakes can be close to unity, so
most lines of sight pass through a turbulent region. This
can help to reconcile the available evidence of turbulence in such clusters as
Perseus with the existence of long filaments in the
intra\-cluster gas that seem to be inconsistent with pervasive turbulence.

Since typical lines of sight through a cluster pass through a small number
of turbulent cells, synchrotron emission from galaxy clusters can show
a detectable degree of polarization at wavelengths 3--6 cm.

%-------------------------------------------------------------------------
\section{Three evolutionary stages}

\subsection{The epoch of major mergers}
Theories of hierarchical structure formation suggest that clusters of galaxies
have been assembled relatively recently. $N$-body simulations indicate that
the clusters form at the intersection of dark matter filaments in the
large-scale structure, and result from both major mergers of objects of
comparable mass (of order $10^{15}M_\odot$) and  the accretion of smaller
clumps onto massive protoclusters. It is likely that intense random vortical
flows, if not turbulence, are produced in the merger events.
Their
plausible
properties are summarized in Table~\ref{res}. The structure of magnetic field
at this stage is illustrated in the left-hand panel of Fig.~\ref{snapshots}.
What is shown is the statistically steady state of magnetic field produced by
dynamo action in turbulent flow
with the Reynolds number about 400 and the magnetic Prandtl number equal to
unity. Similar
magnetic structure plausibly occur
in the turbulent wakes of subclusters and galaxies as well.

%------------------------------------------------------------------------
\begin{table*}%[h]
\caption{Summary of turbulence and magnetic field parameters at
various stages of cluster evolution: duration of the stage (the last two
stages represent steady states), the r.m.s.\ velocity $v_0$ and scale $l_0$
of turbulence and eddy turnover time $t_0=l_0/v_0$ (for the decaying turbulence,
values for the middle of the decay stage are given, 2\,Gyr after its start),
the equipartition magnetic
field $B_\mathrm{eq}=(4\pi\rho v_0^2)^{1/2}$ with $\rho$ the gas density
(i.e., maximum field strength within a turbulent cell),
thickness of magnetic filaments and sheets $l_B$ for the statistically steady
state of the dynamo, the r.m.s.\
magnetic field within a turbulent cell $B_\mathrm{rms}$,
and
finally the standard deviation of the Faraday rotation measure $\sigma_\RM$
(calculated
 for the volume filling turbulence
along path length of $750\kpc$ through the central parts of a cluster
in the first two lines, and
assuming one transverse wake along the line of sight in the last two lines).
A subcluster mass of $3\times10^{13}M_\odot$ has been assumed.}
\label{res}
\begin{center}
\begin{tabular}{lccccccccc}\hline
Evolution stage &Duration&$v_0$     &$l_0$    &$t_0$     &$B_\mathrm{eq}$&$l_B$   &$B_\mathrm{rms}$ &$\sigma_\RM$\\
                &(Gyr)   &($\!\kms)$&(kpc)    &(Gyr)     &($\mu$G)       &(kpc)   &($\mu$G)               &($\!\radm$)\\
\hline
Major mergers   &4       &300       &150      &0.5\phzero&4              &25      &1.8                    &200\\
Decaying turbulence
                &5       &130       &260      &2.0\phzero&2              &44      &0.8                    &120\\
Subcluster wakes&        &260       &200      &0.8\phzero&4              &34      &1.6                    &110\\
Galactic wakes  &        &300       &\tphzero8&0.03      &4    &\phzero\tphzero1.4&1.6                    &\tphzero5\\
\hline
\end{tabular}
\end{center}
\end{table*}
%------------------------------------------------------------------------

It is not quite clear whether
random flows driven during major merger
events and at later stages of evolution will develop into turbulence. The
nature of the flow depends on the value the Reynolds number which is difficult
to estimate reliably for the collisionless, magnetized plasma of the
intra\-cluster space where plasma instabilities can be responsible for anomalous
viscosity and resistivity (Schekochihin et al.\ 2005). The problem is further
complicated by the possibility of dynamo action,
since the magnetic
field can affect both viscosity and magnetic diffusivity.
This may lead to the growth of the magnetic diffusivity and reduction of
viscosity as the magnetic field is being amplified by the dynamo,
so that the magnetic Prandtl number tends to unity.

\subsection{Decaying turbulence}

Random flows produced by major mergers decay
after the end of the merger event. Unlike a laminar flow that decays
exponentially in time due to viscosity, turbulent kinetic energy decays
slower, as a power law (e.g., Landau \& Lifshitz 1975; Frisch 1995). The
reason for this is that kinetic energy mainly decays at small scales, to where
it is constantly supplied by the turbulent cascade. As a result, the energy
decay rate depends nonlinearly on the energy itself, which makes the decay a
power law in time. Our simulations confirm that the power-law decay occurs
even for the Reynolds number as small as $\Rey\approx100$. At this stage of
evolution, the turbulent scale $l_0$ grows with time, whereas turbulent
energy density $E$ reduces, together with the turbulent speed $v_0$,
typically as
\[
E\simeq \sfrac{1}{2}v_0^2\propto (t/t_0\ini)^{-6/5},
\quad
l_0\propto (t/t_0\ini)^{2/5} \quad\mbox{for }t/t_0\ini\gg1\;,
\]
where subscript `i' refers to the start of the evolution, $t_0\ini$ is a
certain dynamical time scale, which can be identified with the initial
turnover time of the energy-containing eddies, $t_0\ini=l_0\ini/v_0\ini$,
subscript `0' refers to the
energy-range (correlation) scale of the motion.
The
structure of magnetic field in the decaying flow is shown in the right-hand
panel of Fig.~\ref{snapshots}, and parameters of the flow and magnetic field
are shown in the second line of Table~\ref{res}.

%------------------------------------------------------------
\subsection{Turbulent wakes of subclusters and galaxies}
At the final stage of the evolution, when the cluster enters a steady state,
turbulence is maintained only in the wakes of galaxies and smaller mass clumps
that continue to accrete onto the cluster. The wakes become weaker as the gas
within the clumps or galaxies is stripped by the ram pressure of intra\-cluster
gas. The radius of a wake at its head is close to the radius within which gas
of the mass clump or galaxy remains intact. We estimate the stripping radius
as $R_0\simeq100\kpc$ for clumps of a mass $10^{13}M_\odot$ (which fall into a
cluster every 3 Gyr) and $R_0=3$--$5\kpc$ for massive elliptical galaxies. If
the flow within the wake becomes turbulent (so that it can be described in
terms of Prandtl's theory of turbulent wakes), the wake length $X$ is
controlled by the magnitude of the Reynolds number via
\[
X/R_0\simeq \left(\Rey\ini/\Reycr\right)^3\;,
\]
where $\Reycr
\approx
400$ (Tomboulides \& Orszag 2000) is the marginal Reynolds
number with respect to the onset of turbulence.
This value of $\Reycr$ was obtained for a flow around a solid sphere; $\Reycr$
for gas spheres is not known.
 The strong dependence of the
wake parameters on the Reynolds number makes the estimates somewhat uncertain.
On the other hand, it implies that galactic wakes can be very sensitive to the
detailed parameters of the galactic motion and intergalactic gas, so that
clusters with very similar parameters can
have vastly different wake structures.

The area covering and volume filling factors, $f_S$ and $f_V$, respectively,
of $N=5$ wakes, produced by  $10^{13}M_\odot$ subclusters,
within the
virial
radius $r\approx3\Mpc$, are estimated as
\[
f_S\simeq 0.15\,\frac{N}{5}\,\left(\frac{R_0}{100\kpc}\right)^6
\left(\frac{\Reycr}{400}\right)^{-4}
\left(\frac{\tilde\lambda}{1\kpc}\right)^{-4},
\]
\[
f_V\simeq 0.02\,
\frac{N}{5}\,\left(\frac{R_0}{100\kpc}\right)^8
\left(\frac{\Reycr}{400}\right)^{-5}
\left(\frac{\tilde\lambda}{1 {\rm kpc}}\right)^{-5},
\]
where $N\approx5$ is consistent with models of hierarchical structure formation,
and $\tilde\lambda$ is an effective mean free path in the intra\-cluster
gas (its introduction is an attempt to allow for our insufficient
understanding of viscosity mechanisms). The covering and filling factors strongly depend on
$\tilde\lambda$ and $\Reycr$.  Furthermore,
both $f_S$ and $f_V$ depend on high powers of another poorly known parameter,
the stripping radius $R_0$. Hence,
as noted above,
 properties of the subcluster wakes can be
rather different in apparently similar clusters. In addition,
numerical simulations of turbulent wakes should be treated with caution as
otherwise reasonable approximations, numerical resolution, and numerical
viscosities can strongly affect the results. On the other hand, it is plausible
that $f_S=O(1)$ but $f_V\ll1$, so that a typical line of sight through the
cluster intersects at least one turbulent region (where our estimate of the
r.m.s.\ turbulent speed is $200$--$300\kms$) despite the fact that turbulence
occurs only in a small fraction of the cluster volume.
Thus the presence of long H$\alpha$
filaments observed by Fabian et al.\ (2003, 2006) in the core of the Perseus
cluster may not be inconsistent with various evidence for random motions
in this cluster core (Churazov et al.\ 2004; Rebusco et al.\ 2006).
A possible signature of such spatially intermittent turbulence  could be a
specific shape of spectral lines, with a narrow core, produced in quiescent
regions, accompanied by nonthermally broadened wings.

The area covering factor of galactic wakes within the gas core radius, 180
kpc, is unity if
\begin{equation}\label{XR0}
X/R_0\simeq 30\mbox{--}15\;,
\quad
X\simeq 100\mbox{--}70\kpc\;,
\end{equation}
and the volume filling factor of such wakes is $f_V\simeq0.07$.
The
length
 of galactic wakes required to cover the projected cluster area, given
by Eq.~(\ref{XR0}), does not seem to be unrealistic. For example, Sakelliou et
al.\ (2005) have observed a wake behind a massive elliptical galaxy (mass of
order $2\times10^{12}M_\odot$) moving through the intra\-cluster gas at a speed
about $v_\mathrm{c}\simeq1000\kms$. The length of the detectable wake is about
$X\simeq130\kpc$ (assuming that it lies in the sky plane), and its mean radius
is $40\kpc$ (obtained from the quoted volume of about $2\times10^{6}\kpc^3$).
The projected area of the wake is about $10^4\kpc^2$, as compared to
$10^3\kpc^2$ for the wake parameters derived above. This wake has been
detected only because it is exceptionally strong, and it is not implausible
that weaker but more numerous galactic wakes can cover the
projected
area of the central
parts of galaxy clusters.

We conclude that subcluster wakes are likely to be turbulent, but galactic
wakes can be laminar if the viscosity of the intra\-cluster gas is as large as
Spitzer's value. Given the uncertainty of the physical nature (and hence,
estimates) of the viscosity of the magnetized intra\-cluster plasma, we suggest
that turbulent galactic wakes remain a viable possibility. Both types of wake
have low volume filling factor but can have an area covering factor of order
unity. Parameters of turbulence and magnetic fields produced within the wakes
are given in
the last two lines of
 Table~\ref{res}.

%-------------------------------------------------------------------------
\section{Intracluster magnetic fields and their observational signatures}
A random flow of electrically conducting fluid (either turbulent or not) can
generate and maintain magnetic field if the magnetic Reynolds number
$\Rm=v_0l_0/\eta$ exceeds a certain critical value, $\Rm>\Rmcr$ where
$\Rmcr$ depends on the statistical properties of the flow and usually
remains within the range $\Rmcr=30$--$100$ (here $\eta$ is the magnetic
diffusivity). This condition is undoubtedly satisfied in the
intra\-cluster plasma, where $\Rm\gg1$ (unlike the kinematic Reynolds number
$\Rey$ which may be quite modest), and so the dynamo action in galaxy clusters
is more than plausible. This type of dynamo, known as the fluctuation dynamo
because it produces fluctuating magnetic fields with zero mean value, does not
require any overall rotation, density stratification, $\alpha$-effect, etc.
Reviews of fluctuation dynamos can be found in Zeldovich et al.\
(1990) and Brandenburg \& Subramanian (2005).

At an early stage of the dynamo action, when magnetic fields are still
too weak to affect the flow, magnetic energy density grows exponentially.  The
e-folding time of the r.m.s.\ magnetic field generated by motions of a scale
$l$ and speed $v$ is of the order of
$\tau\simeq l/v$. In Kolmogorov turbulence (and in any random flow with a sufficiently
steep energy spectrum), the e-folding time is shorter at smaller scales, so
that magnetic energy spectrum has maximum at small scales, close to the
magnetic diffusion scale
$l_\eta=l_0\Rm^{-d}$ with $d=1/2$
(Kazantsev 1967;
Zeldovich et al.\ 1990; Schekochihin et al.\ 2004)
or $d=3/4$
for the marginal mode in Kolmogorov turbulence
(Brandenburg \& Subramanian 2005 and references therein).
Thus, magnetic field is
concentrated into magnetic sheets
(and filaments)
of thickness $l_\eta$ and
length of order $l_0$.
The field strength within the magnetic structures is,
at saturation,
close to
equipartition with kinetic energy, $B\simeq B_\mathrm{eq}=(4\pi\rho
v_0^2)^{1/2}$, where $\rho$ is the gas density. An application of this theory
to galaxy clusters is discussed by Ruzmaikin et al.\ (1989), although these
authors presumed that
turbulence produced in galactic wakes can be volume-filling.

The nonlinear stage of the fluctuation dynamo, where the growth of magnetic
energy saturates because of the action of the Lorentz force on the flow, is
more controversial (Brandenburg \& Subramanian 2005). We consider plausible a
model of the nonlinear dynamo suggested by Subramanian (1999); this model is
consistent with numerical simulations of the fluctuation dynamo, especially for
systems with unit magnetic Prandtl number (Subramanian
et al.\ 2006 and references therein). In this model, the thickness of magnetic structures
in the saturated dynamo is of the order of the magnetic diffusion
scale, but now based on the {\em critical\/} value of the magnetic
Reynolds number,
\[
l_B\simeq l_0 \Rmcr^{-1/2}\;.
\]
This scale is given in Table~\ref{res} for various stages of
evolution, together with the magnetic field within magnetic structures,
$B_\mathrm{eq}$ and the resulting r.m.s.\ magnetic field within a turbulence
cell, $B_\mathrm{rms}$. Table~\ref{res} also contains estimates of the
r.m.s.\ value of the Faraday rotation measure, $\sigma_\mathrm{RM}$ produced
in the dynamo-generated magnetic field; the estimates agree with observations
fairly well. Our numerical simulations of the fluctuation dynamo, where we
directly measure $\sigma_\mathrm{RM}$, also
confirm our semi-analytical estimates.

Our model implies that the correlation scale of random motions in the
intra\-cluster gas, $l_0$, is larger than that assumed earlier. With
$l_0\simeq150\kpc$ (Table~\ref{res}), only 5 turbulent cells occur along a
path length of $L=750\kpc$. The resulting degree of polarization of radio
emission from clusters with synchrotron halos can be estimated as
$p\simeq\frac{1}{2}p_0/n^{1/2}\simeq0.2$, where $p_0\approx0.7$,  $n\simeq L/l_0$
is the number of magnetic structures along the line of sight (assuming that one
magnetic sheet with well-ordered magnetic field occurs in each turbulent
cell and that the linear resolution is better than $l_0$), and a factor $1/2$ allows, in a very approximate manner, for the
volume-filling magnetic field outside the magnetic sheet which only produces
unpolarized emission. Depolarization by Faraday dispersion and beam
depolarization can reduce the degree of polarization to a fraction of percent
at long wavelengths. However, polarization observations at
wavelengths 3--6 cm (where Faraday depolarization is sufficiently weak)
can reveal magnetic structures produced by the dynamo
action if the angular resolution is high enough.
Such observations can become feasible for many galaxy clusters with the advent
of the SKA (Feretti \& Johnston-Hollitt 2004).

\end{document}